\documentclass{elsart}

\usepackage[english]{babel}
\usepackage{amssymb}
\usepackage{amsmath}
\usepackage{psfrag}
\usepackage{epsfig}
\usepackage{graphicx}
\usepackage{graphics}
\usepackage{color}

\begin{document}

\begin{frontmatter}

\title{The Network of Scientific Collaborations within the European Framework Programme}

\date{\today}

\author[urjc1]{Juan A. Almendral}
\author[aveiro]{J. G. Oliveira}
\author[urjc2]{L. L\'{o}pez}
\author[aveiro]{J. F. F. Mendes}
\author[urjc1]{Miguel A. F. Sanju\'{a}n}

\address[urjc1]{Departamento de F\'{\i}sica, Universidad Rey Juan Carlos, Tulip\'{a}n s/n, 28933 M\'{o}stoles, Madrid,
Spain.}
\address[aveiro]{Departamento de F\'{\i}sica, Universidade de Aveiro, Campus Universit\'ario de Santiago, 3810-193
Aveiro, Portugal}
\address[urjc2]{Departamento de Ingeniería Telem\'{a}tica y Tecnolog\'{\i}a Electr\'{o}nica, Universidad Rey Juan Carlos,
Tulip\'{a}n s/n, 28933 M\'{o}stoles, Madrid, Spain.}

\maketitle

\begin{abstract}
We use the emergent field of Complex Networks to analyze the network of scientific collaborations between entities
(universities, research organizations, industry related companies,...) which collaborate in the context of the
so-called Framework Programme. We demonstrate here that it is a scale--free network with an accelerated growth, which
implies that the creation of new collaborations is encouraged. Moreover, these collaborations possess hierarchical
modularity. Likewise, we find that the information flow depends on the size of the participants but not on geographical
constraints.
\end{abstract}


\end{frontmatter}

\section{Introduction}

The interplay between the flow of information in a system and its structure is an open question whose analysis demands
new techniques. To study this issue, it can be used the emergent field in physics research referred to as complex
networks~\cite{albert02}. Many other real systems, such as Internet, biological or social networks, are now best
understood from this point of view. This is indeed the reason why investigations on this subject have been attracting
so much attention in the past few years~\cite{guimera05,jeong00,mendes03}.

The success of complex networks is based on a novel approach to Nature. Namely, the comprehension of a real complex
system cannot be reduced to the study of its constituent elements, i.e., it is necessary a complete analysis of the
relations among all its components. This can be done in two main ways, either proposing theoretical models or
investigating real systems. This paper corresponds to the latter case since we focus our attention on a network of
scientific collaborations in Europe, the so-called Framework Programme---a set of initiatives which define the
priorities for the European Union's research and technological development. We choose the Framework Programme (FP)
because it is a system where structure and information flow affect each other simultaneously, which is interesting
since it is usual to find that either the topology of a network constrains the flow of information on
it~\cite{newman04sci} or the information stored in the network defines its topology~\cite{newman03rev}.

We demonstrate here that the FP is a scale--free network~\cite{barabasi99} with an accelerated growth, which implies
that some form of synergy encourages the creation of new collaborations~\cite{almendral07}. These collaborations possess hierarchical
modularity implying, in this case, that the information flow depends on the size of the participants but not on
geographical constraints~\cite{barabasi03}.

The outline of the paper is as follows. In Sec.~\ref{main_fp5} we present the main features of the FP5 (degree
distribution, clustering coefficient, shortest path distribution and the degree--degree correlation), focusing our
attention only on the technical details. We leave the discussion of the results and how they are related for
Sec.~\ref{disc}.

\section{Main features of the Fifth Framework Programme \label{main_fp5}}

In order to analyze a completely finished programme, we focused our investigation in the Fifth Framework Programme
(FP5) corresponding to the period $1998-2002$. It consists in a set of projects whose participants are, basically,
companies devoted to industrial or commercial business and scientific or educational institutions.

A network is just a set of entities interacting among each other, following certain topology. This can be rendered as a
graph where the elements are represented by a set of points, called nodes or vertices, and the interactions are
regarded as a set of lines between them, called edges or links. Thus, the first step to analyze a network is to fix the
vertices and the property determining if there exists a connection between any couple of them. In our case, it is
natural to consider that each vertex is a participant in the programme and each edge represents two participants
collaborating in a project.

Once the vertices and the edges of the network are defined, the data to generate the graph can be obtained from
CORDIS~\cite{cordis}. This information is not given in the form of a database, thus it is necessary to program a robot
to gather it. The result is a large database made of $15,\!776$ projects, from which it is derived a graph with
$25,\!287$ nodes (participants) and $329,\!636$ edges (collaborations).

To characterize the FP5, we will compute four important features in any network: degree distribution, clustering
coefficient, shortest path distribution and the degree--degree correlation.

\subsection{Degree distribution}

\begin{figure}[htb]
\psfrag{all}{} \psfrag{rall}{} \psfrag{o}{\scriptsize{o}} \psfrag{0}{\scriptsize{0}} \psfrag{1}{\scriptsize{1}}
\psfrag{2}{\scriptsize{2}} \psfrag{3}{\scriptsize{3}} \psfrag{-4}{\scriptsize{-4}} \psfrag{-3}{\scriptsize{-3}}
\psfrag{-2}{\scriptsize{-2}} \psfrag{-1}{\scriptsize{-1}} \psfrag{logk}{log k} \psfrag{logpk}{log P(k)}

\centering
\includegraphics[width=.7\textwidth]{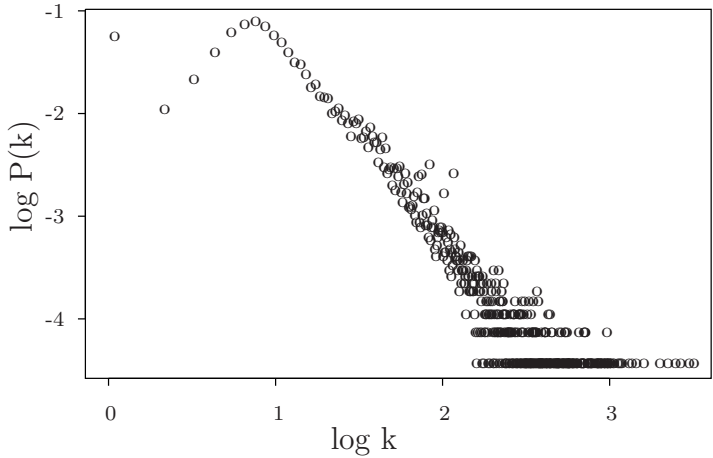}

\caption{This figure depicts the degree distribution of the FP5 network on a $\log - \log$ scale. The average degree is
$k =26.1$ and the maximum degree is a striking $k_{max}=2,784$. The points lying on the right of the maximum fit well a
power--law, $P(k) \sim k^{-\gamma}$, where $\gamma=1.86 \pm 0.02$ with coefficient of determination $R^2=0.85$.}
\label{F:P(k)}
\end{figure}

The {\it degree} of a vertex $i$, $k_i$, is defined as the number of edges which are connected to $i$. We can then
calculate the degree distribution $P(k)$, which gives us the probability of finding a vertex with degree $k$. We find
that the degree distribution of the FP5 follows a power--law, $P(k) \sim k^{-\gamma}$, with a striking maximum degree
$k_{max}=2,\!784$ and average degree $\overline{k} = 26.1$. The distribution can be seen in Fig.~\ref{F:P(k)}, where
the Y axis is $\log P(k)$ and the X axis $\log k$.

Note that, in general, the observation of a power--law is troubling because it may be hindered by the fluctuations at
large degrees~\cite{mendes02}. To measure the degree distribution confidently, it is required that $N \gtrsim 10^3$
when $\gamma < 2$ or $N \gtrsim 10^{2.5(\gamma -1)}$ when $\gamma >2$. Consequently, the scale--free behavior in the
FP5 can only be assured if the points are fitted to a power--law with $\gamma <2.7$.

If a standardized major axis (SMA) regression is calculated for all the points lying on the right of the maximum, they
fit a power--law $P(k) \sim k^{-\gamma}$ with $\gamma =1.86 \pm 0.02$ and coefficient of determination $R^2=0.85$.
Although it is simpler to compute $\gamma$ by means of a linear regression on log-binned data (yielding $\gamma =2.1$),
we choose the SMA method because the result is unbiased~\cite{newman05} and this is important for the discussion in
Sec.~\ref{disc}. Nevertheless, the main result is not the concrete value of $\gamma$ since, as a consequence of the
finite size of a network, there is always a cutoff region which makes difficult to derive it accurately. The main
result is that the degree distribution is described by a power--law with $\gamma < 2$.

\subsection{Clustering coefficient}

The {\it local clustering} of a vertex $i$, $C_i$, is defined as the ratio between the number $y$ of edges connecting
the $k_i$ nearest neighbors of $i$ and the total number of possible edges between these nearest neighbors:
\[ C_i = \frac{2y}{k_i(k_i-1)}. \]
Notice that $C_i$ is only defined for those vertices $i$ that have degree greater than $1$.

It is found that $16,\!313$ of the vertices in the FP5 have $C_i=1$, indicating the presence of many completely
connected clusters. This is due to the fact that $15,\!814$ of these entities participate only in one project, having
as neighbors other vertices, which in turn are all connected between them by virtue of the participation in the
project.

The {\it clustering coefficient} of a network, $\overline{C}$, is just the average value of $C_i$. The FP5 network has
$\overline{C} =0.852$, which is much higher than the clustering coefficient of an Erd\"os--R\'enyi graph with the same
$N$ and $\overline{k}$. Actually, this random graph has $\overline{C} \cong \overline{k}/N$, which is $3$ orders of
magnitude smaller~\cite{bollobas85}.

\begin{figure}[htb]
\psfrag{all}{} \psfrag{rall}{} \psfrag{o}{\scriptsize{o}} \psfrag{0.0}{\scriptsize{0.0}} \psfrag{0.5}{\scriptsize{0.5}}
\psfrag{1.0}{\scriptsize{1.0}} \psfrag{1.5}{\scriptsize{1.5}} \psfrag{2.0}{\scriptsize{2.0}}
\psfrag{2.5}{\scriptsize{2.5}} \psfrag{3.0}{\scriptsize{3.0}} \psfrag{3.5}{\scriptsize{3.5}}
\psfrag{-0.5}{\scriptsize{-0.5}} \psfrag{-1.0}{\scriptsize{-1.0}} \psfrag{-1.5}{\scriptsize{-1.5}} \psfrag{logk}{log k}
\psfrag{logck}{log C(k)}

\centering
\includegraphics[width=.7\textwidth]{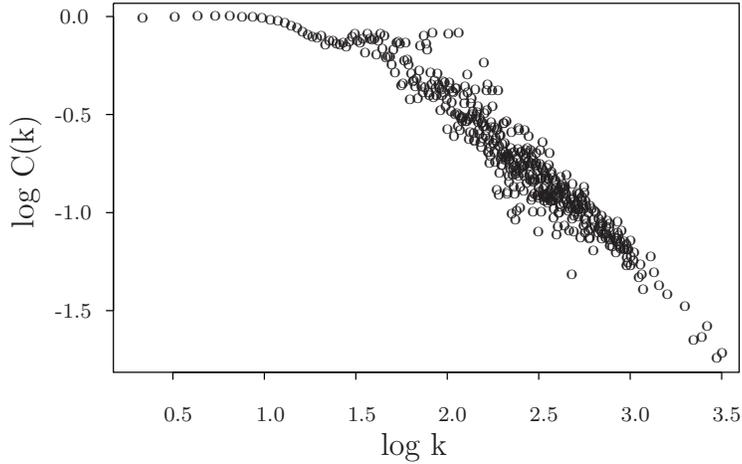}

\caption{In this figure the clustering coefficient as a function of $k$ in a $\log - \log$ plot is shown. After the
initial plateau, where $C(k)$ is approximately constant, it decays as a power--law, $C(k) \sim k^{-\alpha}$, where
$\alpha =0.77 \pm 0.01$ with $R^2=0.88$.} \label{F:C(k)}
\end{figure}

Also, we have measured the clustering coefficient as a function of the degree $k$. To obtain $C(k)$, we consider all
vertices with degree $k$ and, for these vertices, compute the average value. The function $C(k)$ for the graph can be
regarded in Fig.~\ref{F:C(k)}, where the X and Y axes represent $\log k$ and $\log C(k)$ respectively. As it can be
seen, after the region of low values of $k$, where $C(k)$ is approximately constant, it decays as a power--law of $k$.
Thus, if the initial plateau is not considered, the FP5 network verifies that $C(k) \sim k^{-\alpha}$, where $ \alpha
=0.77 \pm 0.01$ with coefficient of determination $R^2=0.88$.

\subsection{Shortest paths}

A {\it path} in a graph is defined as a sequence of vertices in which each successive vertex, after the first, is
adjacent to its predecessor in the path. In unweighted graphs, all edges have the same weight, namely one. The weight
of a path is just the sum of the weights of its edges.

A path between a given pair of vertices is said to be a {\it shortest path} if its weight is minimal. Then the {\it
distance} $\ell_{ij}$ between vertices $i$ and $j$ is defined as the weight of the shortest path that connects these
two vertices. Thus, for unweighted graphs this is just the number of edges of that shortest path.

Notice that in the former paragraph it is assumed that the path between two nodes exists. However, in general, it is
not always possible to define a path between every pair of nodes, and when this happens the graph is made of two or
more connected components. The connected component with more nodes is referred to as the {\it largest connected
component} (LCC).

The FP5 network is not a completely connected component, but we find that the LCC spans $91.17\%$ of the nodes
($23,\!055$ vertices). Hence, we can focus our study only in the largest component since the bulk of the network
belongs to it.

\subsubsection{Distribution of shortest paths}

\begin{figure}[htb]
\psfrag{rall}{} \psfrag{o}{\scriptsize{o}} \psfrag{0.0}{\scriptsize{0.0}} \psfrag{0.1}{\scriptsize{0.1}}
\psfrag{0.2}{\scriptsize{0.2}} \psfrag{0.3}{\scriptsize{0.3}} \psfrag{0.4}{\scriptsize{0.4}}
\psfrag{0.5}{\scriptsize{0.5}} \psfrag{0.6}{\scriptsize{0.6}} \psfrag{1}{\scriptsize{1}} \psfrag{2}{\scriptsize{2}}
\psfrag{3}{\scriptsize{3}} \psfrag{4}{\scriptsize{4}} \psfrag{5}{\scriptsize{5}} \psfrag{6}{\scriptsize{6}}
\psfrag{7}{\scriptsize{7}} \psfrag{8}{\scriptsize{8}} \psfrag{9}{\scriptsize{9}} \psfrag{pl}{P($\ell$)}
\psfrag{l}{$\ell$}

\centering
\includegraphics[width=.7\textwidth]{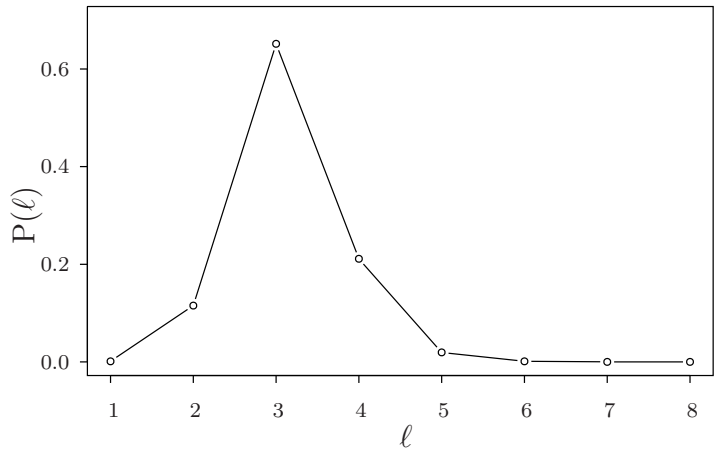}

\caption{The distance distribution $P(\ell)$ in the largest connected component of the FP5 is shown. The mean value is
$3.14$ and the farthest pair of nodes in the graph is separated by only $8$ edges.} \label{F:P(l)}
\end{figure}

We have obtained the distance distribution $P(\ell)$ and the average distance $\overline{\ell}$ for the LCC of the
networks in study. In Fig.~\ref{F:P(l)}, we plot the distance distribution $P(\ell)$ versus $\ell$ to show that the FP5
network displays the small--world effect~\cite{strogatz01}. The greatest distance in the network is only $8$ and the
average distance is $\overline{\ell} =3.14$, which is approximately the value obtained for a random graph with the same
$N$ and $\overline{k}$, i.e., $\overline{\ell} \approx \ln N/\ln \overline{k} = 3.11$.

\subsubsection{$\overline{\ell}$ as a function of $k$}

It is also possible to calculate the average distance of a vertex of degree $k$ to all other vertices in the LCC. To
obtain $\overline{\ell}(k)$, one first calculates the average distance from vertex $i$, $\overline{\ell}_i$, to all
other vertices in the LCC and then averages over all vertices $i$ which have $k_i = k$.

\begin{figure}[htb]
\psfrag{all}{} \psfrag{rall}{} \psfrag{o}{\scriptsize{o}} \psfrag{0.0}{\scriptsize{0.0}} \psfrag{0.5}{\scriptsize{0.5}}
\psfrag{1.0}{\scriptsize{1.0}} \psfrag{1.5}{\scriptsize{1.5}} \psfrag{2.0}{\scriptsize{2.0}}
\psfrag{2.5}{\scriptsize{2.5}} \psfrag{3.0}{\scriptsize{3.0}} \psfrag{1.5}{\scriptsize{1.5}}
\psfrag{2.0}{\scriptsize{2.0}} \psfrag{2.5}{\scriptsize{2.5}} \psfrag{3.0}{\scriptsize{3.0}}
\psfrag{3.5}{\scriptsize{3.5}} \psfrag{4.0}{\scriptsize{4.0}} \psfrag{logk}{log k} \psfrag{lk}{$\overline{\ell}(k)$}

\centering
\includegraphics[width=.7\textwidth]{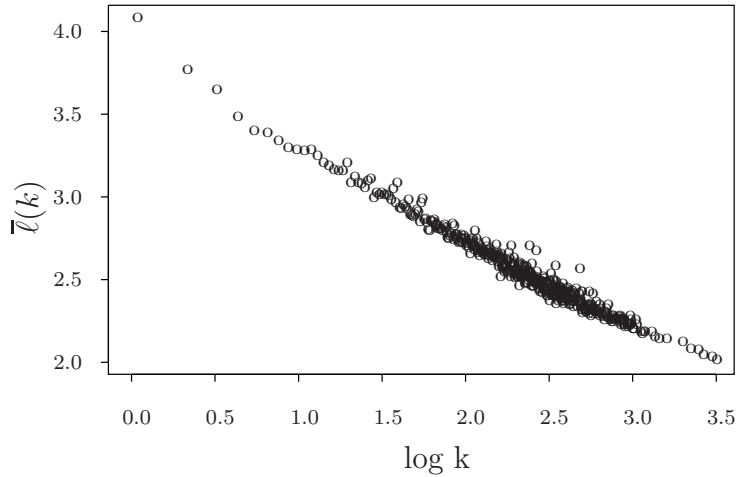}

\caption{The average distance of a vertex of degree $k$ to all other vertices in the LCC is depicted. It can be seen
the logarithmic dependence of the points since it is verified that $\overline{\ell}(k) \sim \log k^{-\beta}$ where
$\beta =0.555 \pm 0.004$ with $R^2=0.97$.} \label{F:l(k)}
\end{figure}

In Fig.~\ref{F:l(k)} we plot $\overline{\ell}(k)$ for the LCC. It is a figure in a linear--log scale, where the Y axis
means $\overline{\ell}(k)$ and the X axis is $\log k$. It is verified that $\overline{\ell}(k) \sim \log k^{- \beta}$
where $\beta = 0.555 \pm 0.004$ with $R^2=0.97$.

\subsection{Degree--degree correlation}

An interesting question is which vertices pair up with which others. It may happen that vertices connect randomly, no
matter how different they are. But usually there is a selective linking, there is some feature which makes more (or
less) likely the connection~\cite{newman03rev}. If nodes with the same feature tend to link among them, the situation
is called {\it assortative mixing}. In the opposite case, when vertices with some feature do not tend to connect among
them, we have {\it disassortative mixing}.

A property which is usually used to investigate the presence of assortative mixing is the degree correlation. In this
case, we say that there is assortative mixing when the nearest neighbors of vertices with high degree have also high
degree. And there is disassortative mixing when the nearest neighbors of vertices with high degree have low degree.

To analyze the degree correlations, we carry out three calculations: the joint degree--degree distribution, the mean
degree $\bar{k}_{nn}(k)$ of the nearest neighbors of a vertex of degree $k$ and the assortativity coefficient.

\subsubsection{Joint degree--degree distribution}

The {\em joint degree--degree distribution} $P(k,k')$ gives us the probability of finding an edge which connects
vertices of degree $k$ and $k'$. We measured $P(k,k')$ for the FP5 network and the result for $k<200$ is depicted in
Fig.~\ref{F:P(k,k')}. While the X and Y axes represent the degrees $k$ and $k'$, the Z axis gives the corresponding
probability in per mill.

We see that $P(k,k')$ has sharp peaks for $k=k'$. This means that if one chooses at random a vertex of degree $k$ then,
with great probability, it will be connected to vertices of degree $k=k'$. This result suggests that the FP5 presents
assortative mixing.

\begin{figure}[htb]
\centering
\includegraphics[width=.7\textwidth]{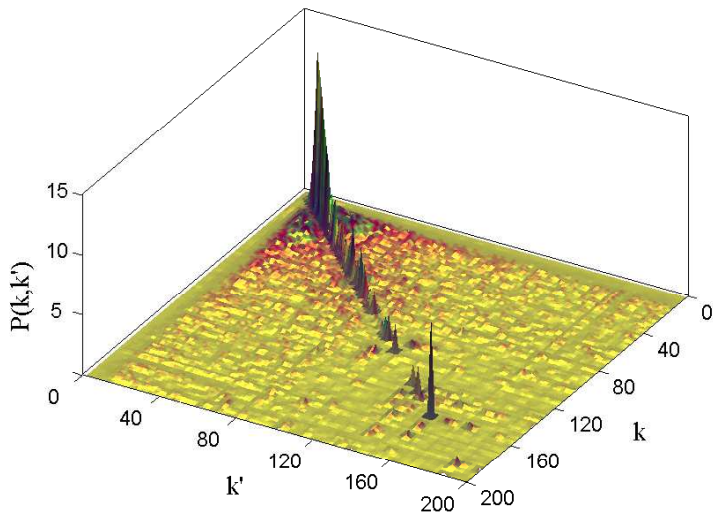}

\caption{Joint degree--degree distribution of the FP5 network. The X and Y axes represent the degrees $k$ and $k'$ and
the Z axis gives the corresponding joint degree-degree probability in per mill. The range is limited from $0$ to $200$
to illustrate a clearer picture. The distribution peaks on the line $k=k'$ which implies that the FP5 shows assortative
mixing.} \label{F:P(k,k')}
\end{figure}

\subsubsection{$\bar{k}_{nn}(k)$ distribution}

It is important to remark that the joint degree--degree distribution requires many points to obtain good statistics.
For example, if we focus our analysis in the range $[0,200]$, we need about $200 \times 200$ points, otherwise
fluctuations are important and the plot is far from smooth~\cite{boguna04}. To avoid this problem, it is used the mean
degree $\bar{k}_{nn}(k)$ of the nearest neighbors of a vertex of degree $k$, which is a coarser but less fluctuating
feature. To compute $\bar{k}_{nn}(k)$ we have only to find all nodes with degree $k$, and then, the average degree of
all their neighbors is calculated.

\begin{figure}[htb]
\psfrag{0}{\scriptsize{0}} \psfrag{1}{\scriptsize{1}} \psfrag{2}{\scriptsize{2}} \psfrag{3}{\scriptsize{3}}
\psfrag{2.0}{\scriptsize{2.0}} \psfrag{2.1}{\scriptsize{2.1}} \psfrag{2.2}{\scriptsize{2.2}}
\psfrag{2.3}{\scriptsize{2.3}} \psfrag{2.4}{\scriptsize{2.4}} \psfrag{2.5}{\scriptsize{2.5}} \psfrag{logk}{log k}
\psfrag{logknn}{$\log \bar{k}_{nn}(k)$} \psfrag{k200}{$k=200$}

\centering
\includegraphics[width=.7\textwidth]{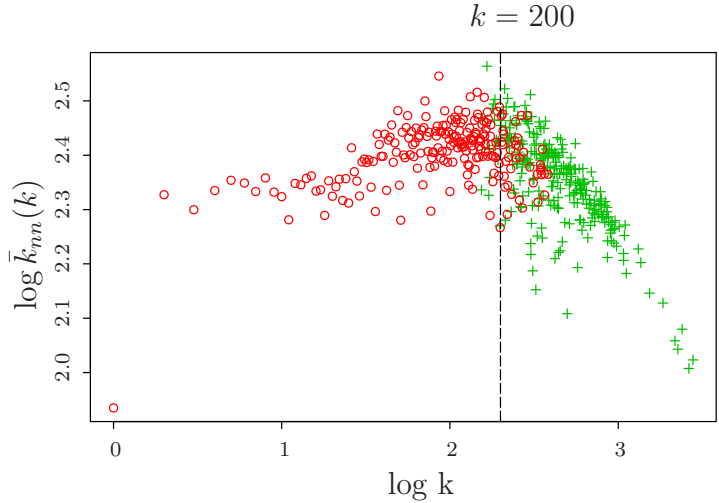}

\caption{Plot of the mean degree of the nearest neighbors of a vertex of degree $k$, $\bar{k}_{nn}(k)$, on a $\log -
\log$ scale. Apparently, the behavior below $k \approx 200$ is assortative and disassortative for higher degrees.
However, only the region with assortative mixing must be considered because over $k \approx 200$ the finite size of the
network is important. If the green crosses are points calculated from only $1$ or $2$ participants and the rest of the
points are red circles, it can be seen that the region with disassortative mixing is essentially made of green
crosses.} \label{F:knn}
\end{figure}

The result is shown in Fig.~\ref{F:knn}, where the X axis represents $\log k$ and the Y axis $\log \bar{k}_{nn}(k)$.
Interestingly, we find that the picture presents two regions with different behaviors which approximately overlap on $k
\approx 200$. While for high degrees ($k \gtrsim 200$) the mixing is disassortative, for low degrees ($k \lesssim 200$)
seems to be assortative. However, the points on the right-hand side correspond to degrees where the finite size of the
network is important, thus we cannot conclude that over $k=200$ there is disassortative mixing.

To show this fact we have represented as green crosses those points calculated from only $1$ or $2$ participants
(indicating the proximity to the cutoff) and the rest of the points as red circles. It can be seen that the majority of
points over $k=200$ are green crosses, that is, the $\bar{k}_{nn}(k)$ obtained for participants with high degrees is
biased by the presence of the cutoff. This is reasonable since participants with $k \approx 1000$ could only have the
value of $\bar{k}_{nn}(k)$ which the tendency imposes, if they had many neighbors with even higher degrees, but the
finite size of the network impedes this.

Then, if we only consider the points below $k \approx 200$ (or equivalently, the red circles), our result suggests that
the mixing is assortative. Nonetheless, the mean degree of the nearest neighbors varies only from $200$ to $316$, thus
another measure of the mixing will be helpful to confirm if the FP5 is assortative.

\subsubsection{Assortativity coefficient}

Another way to quantify the mixing in the FP5 is by means of the {\em assortativity coefficient}~\cite{newman02}. In
this case, we obtain what type of mixing takes place in the network by means of a single number instead of a
distribution.

If $e_{jk}$ is the probability that a randomly chosen edge has vertices with degree $j$ and $k$ at either end, the
assortativity coefficient takes the following form:
\begin{equation*}
r = \frac{\sum_{jk} jk(e_{jk} -q_j q_k)}{\sum_k k^2 q_k - \left( \sum_k k q_k \right)^2}
\end{equation*}
where $q_k = \sum_j e_{jk}$ and $q_j = \sum_k e_{jk}$. This coefficient verifies that $-1 \le r \le 1$, being positive
when the network is assortative and negative when it is disassortative.

We find that the FP5 has assortative mixing because $r=0.04$, which is close to the coefficients obtained for other
social networks~\cite{newman03pre}.

\section{Discussion of the results \label{disc}}

In order to analyze the FP5 as a complex network, we have defined a graph where the vertices are all the participants
and each edge represents two collaborators in (at least) one project. The resulting graph is made of $25,\!287$ nodes
and $329,\!636$ edges.

We have shown that this network is scale--free since its degree distribution follows a power--law, $P(k) \sim
k^{-\gamma}$. Then, we can infer that, during its growth, the collaborations were established by means of some type of
preferential attachment. In other words, the participants with more collaborations establish new ones at higher rate
than participants with few connections. As a consequence, the so-called ``rich--get--richer'' phenomenon arises, in
which the most connected participants increase their collaborations at the expense of the latecomers.

It is interesting to note the fact that $\gamma <2$, since it is known that the average degree of the network diverges
in that case. A possibility to explain this result is that the FP5 is an accelerated growing network. In these
networks, the total number of edges grows faster than a linear function of the total number of vertices and,
consequently, it may be verified that $1< \gamma <2$.

To elucidate this issue, we have computed the average degree during several years to check its tendency. Although we
have only the data corresponding to $4$ years (Table~\ref{table}), they are enough to confirm the existence of an
accelerated growth since $\overline{k}$ is not constant but it grows. But if the collaborations grow faster than
proportional to the number of participants, it is because they do not emerge by the mere increase of participants. Not
only new participants contribute to increase the number of collaborations, but also the old ones. Then, some form of
synergy exists which encourages the creation of new collaborations.

\begin{table}[htb]
\centering
\begin{tabular}{|l|c|c|c|c|}
    \hline Year & 1999 & 2000 & 2001 & 2002 \\
    \hline $N$ & 7,732 & 14,730 & 21,253 & 25,287 \\
    \hline $\overline{k}$ & 16.30 & 19.25 & 23.56 & 26.07 \\
    \hline
\end{tabular}
\caption{This table shows the average degree and the total number of vertices of the FP5 during the four years it
lasted. We can conclude that there is an accelerated growth in the network since the average degree is not constant but
it grows.} \label{table}
\end{table}

We have found that the growth of the mean length of the shortest path between two vertices, $\overline{\ell}(N)$, is
slower than any positive power of $N$. Then, we can state that the FP5 is a network with the small--world effect. This
can be seen easily in the average separation between any two participants, which is approximately $3$ (that is, only
$2$ intermediaries). This compactness is indeed useful to integrate the R+D+I in Europe.

Likewise, the clustering coefficient of the FP5 is much higher than the corresponding to a random graph. Moreover, the
local clustering coefficient depends on the degree as $C(k) \sim k^{-0.77}$. This suggests the existence of a
hierarchical modularity in the FP5 because both scale--free and modular networks are degree--independent, whereas
hierarchical modularity is characterized by the scaling law $C(k) \sim k^{-\alpha}$~\cite{barabasi02}. Therefore, the
FP5 has an inherent self--similar structure, being made of many highly connected small modules (all the participants
with $C_i =1$), which integrate into larger modules, which in turn group into even larger modules~\cite{makse05}.
Furthermore, since this result suggests that the network has weak geographical constraints~\cite{barabasi03}, we
searched for communities in it~\cite{newman04pre} to verify this question and found precisely that they were not based
on nationality.

When we focus our attention in the degree--degree correlations, we find that the FP5 is assortative as it is usual in
social networks. This means that participants with similar degree tend to collaborate more frequently than participants
with different degrees. But we have checked that if a participant has high degree, it is due, in most of the cases, to
being involved in many projects. Then, assuming that nodes with high degree are mainly large institutions, because many
FP projects at the same time require an important support that is more common in large institutions, the assortativity
found in the FP5 means that the collaborations are biased by the size of the institutions.

Since the hierarchical organization of a network is a poorly defined term, we study this question through the notion of
{\it hierarchical path}~\cite{gao01} because it is uncorrelated with $C(k)$. A path is said hierarchical if the degrees
of the vertices along this path vary monotonously or they grow monotonously up to some maximum value, from which
decrease monotonously. Then, the fraction $H$ of shortest paths which are hierarchical can be used as a metric of a
hierarchical topology~\cite{trusina04}. We find that the FP5 has $H=0.91$, which confirms the hierarchical structure of
the FP5. Actually, the distribution of hierarchical shortest paths is rather similar to the distribution of all
shortest paths. And this implies that most of the shortest paths between nodes are hierarchical.

\section{Conclusions}

We have thoroughly analyzed the complex network constituted by the scientific collaborations of the fifth Framework
Programme. The network is scale--free with an accelerated growth, which means that new collaborations are created at a
faster rate than usual. We have also concluded that some sort of synergy among the participants exists since new
collaborations appear. Moreover, we have also found that this network possesses the property of small--world. Due to
the hierarchical modularity property, the FP5 has a self-similar structure and it is also robust to structural changes.
Finally, another important feature is the assortative mixing, which in this case means that collaborations among
participants of similar size appear easier.

\subsection*{Acknowledgements}

JAA, LL, and MAFS acknowledge financial support from MCyT--Spain (BFM 2000-0967 and BFM2003-03081), MEC--Spain
(FIS2006-08525) and the URJC--Spain (URJC-GCO-2003-16). JGO acknowledges financial support from FCT (Portugal) grant
No. SFRH/BD/14168/2003. JFFM was partially supported by Projects POCTI (FAT/46241/2002, MAT/46176/2002) and project
DYSO NET-NEST/012911.


\begin{thebibliography}{9}

\bibitem{albert02} R. Albert and A.-L. Barab\'asi, Rev. Mod. Phys. {\bf 74}, 47--97 (2002).

\bibitem{guimera05} R. Guimer\'a and L.A.N. Amaral, Nature {\bf 433}, 895--900 (2005).

\bibitem{jeong00} H. Jeong, B. Tombor, R. Albert, Z. Oltvai and A.-L. Barab\'asi, Nature {\bf 407}, 651--654 (2000).

\bibitem{mendes03} S.N. Dorogovtsev and J.F.F. Mendes, {\it Evolution of Networks: from Biological Nets to the Internet
and WWW} (Oxford University Press, Oxford, 2003).

\bibitem{newman04sci} J. Balthrop, S. Forrest, M.E.J. Newman and M.M. Williamson, Science {\bf 304}, 527--529 (2004).

\bibitem{newman03rev} M.E.J. Newman, Siam Review {\bf 45(2)}, 167--256 (2003).

\bibitem{barabasi99} A.-L. Barab\'asi and R. Albert, Science {\bf 286}, 509--512 (1999).

\bibitem{almendral07} J. A. Almendral, J. G. Oliveira, L. L\'opez, Miguel A. F. Sanju\'an and Jose F. F. Mendes, New
Journal of Physics (2007). In Press.

\bibitem{barabasi03} E. Ravasz and A.-L. Barab\'asi, Phys. Rev. E {\bf 67}, 026112 (2003).

\bibitem{cordis} Community Research and Development Information Service can be found at http://www.cordis.lu/.

\bibitem{mendes02} S.N. Dorogovtsev and J.F.F. Mendes, Adv. Phys. {\bf 51}, 1079 (2002).

\bibitem{newman05} M. E. J. Newman, Contemporary Physics {\bf 46}, 323-351 (2005).

\bibitem{strogatz01} S.H. Strogatz, Nature {\bf 410}, 268--276 (2001).

\bibitem{bollobas85} B. Bollobas, {\it Random graphs} (Academic Press, London, 1985).

\bibitem{barabasi02} E. Ravasz, A.L. Somera, D.A. Mongru, Z.N. Oltvai and A.-L. Barab\'asi, Science {\bf 297},
1551--1555 (2002).

\bibitem{makse05} S. Chaoming, S. Havlin and H.E. Makse, Nature {\bf 433}, 392--395 (2005).

\bibitem{newman04pre} M.E.J. Newman, Phys. Rev. E {\bf 69}, 066133 (2004)

\bibitem{boguna04} M. Bogu\~n\'a, R. Pastor--Satorras and A. Vespignani, Eur. Phys. J. B. {\bf 38}, 205--209 (2004).

\bibitem{newman02} M.E.J. Newman, Phys. Rev. Lett. {\bf 89}, 208701 (2002).

\bibitem{newman03pre} M. E. J. Newman. Mixing patterns in networks. Phys. Rev. E {\bf 67}, 026126 (2003).

\bibitem{gao01} L. Gao, IEEE/ACM Transactions on networking {\bf 9}, 733 (2001).

\bibitem{trusina04} A. Trusina, S. Maslov, P. Minnhaben and K. Sneppen, Phys. Rev. Lett. {\bf 92}, 178702 (2004).

\end{thebibliography}
\end{document}